\documentclass[prd,superscriptaddress,twocolumn]{revtex4}

\usepackage{caption}
\usepackage{subcaption}
\usepackage{amssymb}
\usepackage{amsfonts}
\usepackage{amsmath}
\usepackage{graphicx}
\usepackage{dcolumn}
\usepackage{bm}
\usepackage[colorlinks]{hyperref}

\begin{document}
\title{Shedding Light on Naked Singularities}

\author{Franco Fiorini}
\email{franco.fiorini@ib.edu.ar}
\affiliation{Grupo de Comunicaciones Ópticas, Departamento de Ingeniería en Telecomunicaciones, Consejo Nacional de Investigaciones Científicas y Técnicas (CONICET) and Instituto Balseiro (UNCUYO), Centro Atómico Bariloche, Av.~Ezequiel Bustillo 9500, CP8400, S. C. de Bariloche, Río Negro, Argentina}
\author{Santiago M. Hernandez}
\email{shernandez@ib.edu.ar}
\affiliation{Grupo de Comunicaciones Ópticas, Departamento de Ingeniería en Telecomunicaciones, Consejo Nacional de Investigaciones Científicas y Técnicas (CONICET) and Instituto Balseiro (UNCUYO), Centro Atómico Bariloche, Av.~Ezequiel Bustillo 9500, CP8400, S. C. de Bariloche, Río Negro, Argentina}
\author{Juan Manuel Paez}
\email{jmpaez@iafe.uba.ar}
\affiliation{Instituto de Astronomía y Física del Espacio (IAFE, CONICET-UBA), Ciudad Universitaria, Buenos Aires, Argentina.}

\begin{abstract}
Electromagnetic waves propagating in the background provided by a spacetime hosting a strong curvature, naked singularity, are fully studied. The analysis is performed not only in the realm of geometrical optics --which, not surprisingly, proves to be inadequate in the strong-field regime characterizing the vicinity of the singularity-- but also in the physical one in which the field amplitudes must necessarily be incorporated into the description. In addition to the expected divergent outcome with regard to the field amplitudes and power flux as the waves approach the singularity, we found a number of regular (bounded) solutions which seem to coexist with the unbridled effects of the spacetime curvature. In some of them, the singularity operates as a perfect mirror reflecting the surrounding fields. Strikingly, other solutions exhibit a perfectly well behaved, bounded power flux as they propagate towards the singularity, suggesting thus the possibility of having electromagnetic energy transference through it.
\end{abstract}

\maketitle
\section{Introduction} 
Nature seems to be in constant conflict with regard to the potential existence of naked singularities. Cosmic Censorship Hypothesis \cite{Penrose}, \cite{Penrose1}, a tentative scheme to get rid of such eerie manifestations of the extreme faces of the gravitational field, not only remains unproven in its various and different versions \cite{Penrose2}, but it also seems to be compelled to confront several, physically significant counter-examples, see, e.g. \cite{Harada} and \cite{Ong}. Naked singularities are inconvenient because their devastating effects would not be concealed by the presence of event horizons, thus, wreaking havoc with the very notion of predictability. Actually, according to the common lore on the subject, naked singularities might produce uncontrolled bursts of information which ultimately will affect the Cauchy evolution, preventing any prediction on the basis of data fixed at a given partial Cauchy surface. A thorough discussion on the potential observability of these objects can be consulted in the seminal papers \cite{Vir1}-\cite{Vir3}.

However, how is then possible that information --in the form of emerging electromagnetic (EM) fields-- could be produced at the singularity without being destroyed by the very unrestrained spacetime curvature in the first place? Or is it that, by acting as a perfectly expelling object, the singularity behaves as a mirror reflecting all the light present in its surroundings? Yet another possibility is perhaps the most intriguing: is it conceivable that EM signals could be able to cross naked singularities and remain intact? It is our intention here to address these questions of fundamental character by exactly solving Maxwell equations on a curved background representing what can be considered the simplest example of a strong curvature, naked singularity. The goodness of our approach lies in the fact that our results apply beyond the realm of geometrical optics (GO), or equivalently, beyond the light-ray description corresponding to null geodesics. This is important because GO is manifestly non applicable in the strong curvature regime provided by this kind of singularity. Surprisingly, aside from a number of totally expected divergent EM field configurations, several solutions exist which remain bounded as they approach the singularity; actually, they are representative of waves either being reflected by the singularity or crossing from one side to the other, carrying EM power between regions which previously were considered causally disconnected. 

\section{Electrodynamics and naked singularities} 
Recently \cite{nos2}, we have considered the metric  
\begin{equation}\label{mettoy-1}
    ds^2=-x^{-2}dt^2+dx^2+dy^2+dz^2\,,
\end{equation}
where $\bar{x}=(x,y,z)$ are Cartesian, non-dimensional coordinates. It is straightforward to verify that $x=0$ constitutes a curvature singularity not shielded by any event horizon, as the non-vanishing covariant components of the Ricci tensor are $R_{tt}=2\,x^{-4}$ and $R_{xx}=-2\,x^{-2}$, whereas the invariants $ R=g^{\mu\nu}R_{\mu\nu}$, $\mathcal{R}^2=R^{\mu\nu}R_{\mu\nu}$ and $\mathcal{K}=R_{\mu\nu\rho\sigma}R^{\mu\nu\rho\sigma}$ read $ R=-4\,x^{-2}$ and $\mathcal{R}^2=\mathcal{K}=2R^2$. Let us take note that the space has negative scalar curvature, which has a direct impact in the behavior of geodesics as briefly described below.

We judge it important to emphasize that the spacetime described by the metric (\ref{mettoy-1}) does not represent a solution of Einstein field equations powered by any known, physical model of matter-energy, even though the energy-momentum tensor related to this metric has the structure $T_{\mu\nu}=2 x^{-2}diag(0,0,1,1)$, which satisfies both the weak and strong energy conditions. It is nonetheless a nice example, mainly due to its amenable nature, to fully understand how light really behaves in the vicinity of a naked singularity. We refer the reader to \cite{nos2} for a detailed exposition on the causal structure of this space. For the nonce, suffice to say that the naked singularity is reflective, repelling both null and timelike geodesics. This is similar to what occurs in already documented examples in the context of General Relativity, see \cite{Newman}, \cite{Maluf}, as well as in other theories of gravity \cite{Glavan}, \cite{nos3}, where null geodesics are effectively reflected in the high curvature regime so that they cannot access the singularity. As a matter of fact, the causal structure of the metric (\ref{mettoy-1}) is closely related to the one describing the super-extreme ($Q>M$) Reissner-Nordstr\"{o}m spacetime, so we are convinced that our toy model will be useful for studying more physical examples where repulsive naked singularities actually appear. Finally, we should mention that the spacetime is geodesically complete, as all causal geodesics (null and timelike) can be extended to arbitrary values of their affine parameters. In this way, free-falling observers and light rays are not harmed by the curvature singularity. However, not all causal curves are complete, only geodesics are; this can be easily exemplified by considering the non-geodesic curves $x^{\mu}(\tau)$ defined by constant values of $y$ and $z$ having
\begin{equation}\label{curnula}
t(\tau)=\frac{a_{0}^2}{2}\sqrt{1+a^2/a_{0}^2}\, \tau^2+t_{0},\,\,\,\, x(\tau)=a_{0}\tau,
\end{equation}
with constants $a$, $a_{0}$ and $t_{0}$. These curves are null if $a=0$ and timelike if $a\neq0$, and they reach $x=0$ when the generalized affine parameter $\tau\in[0,\tau_{0})$ takes a null value. It is easy to show that all these curves have finite generalized affine parameter length $\ell$ as defined by 
\begin{equation}\label{gapl}
\ell=\int_{0}^{\tau_{0}} \left[\sum_{a=0}^{3}\left(g_{\mu\nu}\frac{d x^{\mu}}{d\tau} E^{\nu}_{a}\right)^2 \right]^{1/2} d\tau\,,
\end{equation}
where $E^{\nu}_{a}$ are the local components of a frame $E_{a}$ (i.e. $E^{\nu}_{a} E^{\mu}_{b}\eta^{ab}=g^{\nu\mu}$), which is parallelly propagated along the curves $x^{\mu}(\tau)$. In the language of Ref. \cite{Schmidt} (see also \cite{HE}), these curves are b-incomplete, meaning that the spacetime is null and timelike b-incomplete. This sort of incompleteness has dramatic consequences on massive particles in non-geodesic motion (e.g., an observer in an accelerated rocket), as well as in light being subjected to optical manipulation, such as in an optical fiber.

We are interested in solving the source-free Maxwell equations in the background provided by the metric (\ref{mettoy-1}), namely 
\begin{equation}
F^{\mu\nu}_{\,\,\,\,\,\,\,;\mu}=0\,,\,\,\,\,(\epsilon^{\mu\nu\rho\sigma}F_{\rho\sigma})_{,\mu}=0\,, \label{cov}
\end{equation}
where $F_{\mu\nu}=A_{\nu,\mu}- A_{\mu,\nu}$, and $A_\mu=(\phi,\bar{A})$ is the 4-potential. Making use of the formalism developed first by Tamm \cite{Tamm} and subsequently worked out by Plebanski \cite{Pleb}, Eqs. (\ref{cov}) can be converted into the standard, flat-space Maxwell system 
\begin{eqnarray}
    &&\bar{\nabla} \cdot \bar{D}=0,\,\,\,\,\,\,\,\bar{\nabla} \times \bar{H}-\partial \bar{D}/\partial t=0,\label{eq:maxwell_mediosfuentescom} \\
   &&\bar{\nabla} \cdot \bar{B}=0,\,\,\,\,\,\,\,\,\,\,\,\,\, \bar{\nabla} \times \bar{E}+\partial \bar{B}/\partial t=0,
     \label{eq:maxwell_medioscom}
\end{eqnarray}
provided the following constitutive relations hold:
\begin{equation}
\bar{D}=\textbf{K}\bar{E}+\bar{\Gamma}\times\bar{H}\,,\,\,\,\,
\bar{B}=\textbf{K}\bar{H}-\bar{\Gamma}\times\bar{E}\,.\label{eq_PT_vector_Bcom}
\end{equation}
Here, the components of the matrix $\textbf{K}$ and vector $\bar{\Gamma}$, are obtained by means of the spacetime metric according to
\begin{equation}
 K_{ij}=-\sqrt{-g}\,g^{ij}/g_{tt}\,,\,\,\,\,\,\,\,\,
\Gamma_m=g_{tm}/g_{tt}\,,\label{defmatk}
\end{equation}
where Latin indexes $i,j,m$ refer to $x,y,z$. The idiosyncrasy behind writing the equations (\ref{cov}) as in (\ref{eq:maxwell_mediosfuentescom}) and (\ref{eq:maxwell_medioscom}), is cemented in the context of so called \emph{optical analogue-gravity} models, see for instance, \cite{leon-phil} and \cite{leon-phi2}. This practice not only provides a more Euclidean, three-dimensional visualization of inherently four dimensional spacetime phenomena \cite{Oleg}-\cite{Gaitas}, but also the possibility of experiencing in the lab some of the frequently elusive effects characterizing the strong field regime of the gravitational field \cite{RAD}. In this way, the study of EM waves in
a given curved spacetime ends up being equivalent to the characterization of waves in a 3D material medium described by the constitutive equations (\ref{eq_PT_vector_Bcom}). 

Let us consider the EM waves
\begin{eqnarray}
\bar{\mathcal{E}}(\bar{x},t)&=&\bar{E}(\bar{x})\exp\big[i\,k_{0}\,(\bar{k}(\bar{x})\cdot\bar{x}- t)\big],\notag\\
\bar{\mathcal{H}}(\bar{x},t)&=&\bar{H}(\bar{x})\exp\big[i\,k_{0}\,( \bar{k}(\bar{x})\cdot\bar{x}- t)\big],\label{expEyH2}
\end{eqnarray}
where $\bar{k}(\bar{x})$ is the spatially-dependent, nondimensional wave vector, $k_{0}$ is the wave number of the monochromatic wave in the flat vacuum (i.e., when $\textbf{K}=\textbf{I}$, $\bar{\Gamma}=\bar{0}$, where $\textbf{I}$ is the $3\times3$ identity matrix), and $\bar{E}(\bar{x}), \bar{H}(\bar{x})$ are the field amplitudes, which also depend on the spatial coordinates. After plugging the ansatz (\ref{expEyH2}) into the curl equations (\ref{eq:maxwell_mediosfuentescom}) and (\ref{eq:maxwell_medioscom}), and using (\ref{eq_PT_vector_Bcom}), we obtain 
\begin{align}
&i\,k_{0}^{-1}\,\bar{\nabla}\times\bar{E}=[\bar{\nabla}(\bar{k}\cdot\bar{x})+\bar{\Gamma}]\times\bar{E}-\textbf{K} \bar{H}\,, \label{eq:rot_E_cuasi_planas}\\
&i\,k_{0}^{-1}\,\bar{\nabla}\times\bar{H}=[\bar{\nabla}(\bar{k}\cdot\bar{x})+\bar{\Gamma}]\times\bar{H}+\textbf{K} \bar{E}\label{eq:rot_H_cuasi_planas}\,.
\end{align}
If we were interested in the ray-tracing techniques associated to GO, the left-hand side of (\ref{eq:rot_E_cuasi_planas}) and (\ref{eq:rot_H_cuasi_planas}) would be negligible, because the fields would vary slowly with respect to $k_{0}$, and also $\bar{\nabla}(\bar{k}\cdot\bar{x})\approx \bar{k}$. In this case, the ray trajectories can be obtained from the canonical equations
\begin{equation}
\bar{\nabla}_{\bar{x}} \mathrm{H}_{am} =-d \bar{k}/dt,\,\,\,\,\,
\bar{\nabla}_{\bar{k}}\mathrm{H}_{am}=d \bar{x}/dt, \label{hamiltonec}
\end{equation}
where $\bar{\nabla}_{\bar{k}}\equiv (\partial/\partial k_x,\partial/\partial k_y,\partial/\partial k_z)$, and the identically zero GO Hamiltonian $\mathrm{H}_{am}$ reads \cite{Hamop},\cite{Mackay1},
\begin{equation}\label{hamilton}
\mathrm{H}_{am}\doteq\det(\textbf{K}) -\bar{p}^{\,\intercal} \textbf{K}\,\bar{p}=0,\,\,\,\,\,\,\, \bar{p}\doteq\bar{k}+\bar{\Gamma}.
\end{equation}
In this context, the canonical equations (\ref{hamiltonec}) can be explicitly written as \cite{nos}
\begin{eqnarray}
d \bar{x}/dt&=&-2 \textbf{K}\bar{p},\label{hamiltoncoorfin}\\
d \bar{k}/dt&=&\bar{p}^{\,\intercal}\Big[ [\textbf{K}_{i}-tr(\textbf{K}^{-1}\textbf{K}_{i})\,\textbf{K}]\,\bar{p}+2\textbf{K}\,\bar{p}_{i}\Big]\hat{e}_{i}\,.\label{hamiltonmomfin}
\end{eqnarray}
Here, $\hat{e}_{i}$, $i:x,y,z$ are the Cartesian unit vectors in $\mathbb{R}^3$, $tr(\textbf{K}^{-1}\textbf{K}_{i})$ is the trace of the matrix $\textbf{K}^{-1}\textbf{K}_{i}$, and $\textbf{K}_{i}$ are three matrices whose components are obtained from the components of $\textbf{K}$ by differentiating with respect to the coordinate $x$, $y$ and $z$, respectively. The same applies to $\bar{p}_{i}=\partial\bar{p}/\partial x_{i}=\partial\bar{\Gamma}/\partial x_{i}$.

Let us conclude this section by examining the propagation of light rays in the analogue medium corresponding to metric (\ref{mettoy-1}), i.e., under the assumption of the GO approximation just presented. Using the definitions (\ref{defmatk}), we get  
\begin{equation}\label{eq:K}
\textbf{K}=|x|\,\textbf{I}\,, \,\,\,\,\, \bar{\Gamma} =\bar{0}\,.
\end{equation}
The GO Hamiltonian, Eq. (\ref{hamilton}), is given by
\begin{equation}
    \mathrm{H}_{am}=|x|\left(x^2-|\bar{k}|^2\right)\,.\label{ham}
\end{equation}
Hence, the dispersion relation $\mathrm{H}_{am}=0$ implies $ x^2=|\bar{k}|^2$, and Hamilton's equations (\ref{hamiltoncoorfin}) and (\ref{hamiltonmomfin}) can be written in components as
\begin{align}
 \dot{x}&=-2|x|k_x\,,\,\,\,\,  \dot{y}=-2|x|k_y\,,\,\,\,\,\,  \dot{z}=-2|x|k_z\,,  \label{eq:x}\\
 \dot{k}_x&=-2\operatorname{sgn}(x)\,x^2\,,\,\,\,\,\,\,\,\dot{k}_y=\dot{k}_z=0,    \label{eq:k1}
\end{align}
where $\operatorname{sgn}(x)$ is the sign function, and we used the fact that $\bar{p}=\bar{k}$ because $\bar{\Gamma}=\bar{0}$. Notice that $k_y$ and $k_z$ are constants of motion (conjugate momenta) associated with the cyclic coordinates $y$ and $z$. Equations~(\ref{eq:x}) and (\ref{eq:k1}) are solved by
\begin{eqnarray}
    x(t)&=&\pm|\Tilde{k}| \csc{(2|\Tilde{k}|\,t+t_{0})}\,,\,\,\,\,\,\Tilde{k}\neq 0,
    \label{eq:sol_x}\\
  x(t)&=&\pm(2\,t+t_{0})^{-1}\,,\,\,\,\,\,\Tilde{k}=0,
\label{eq:sol_x_k_0}
\end{eqnarray}
where $t_{0}$ is an integration constant and $\Tilde{k}\doteq k_{y}^2+k_{z}^2$, thus, $\Tilde{k}=0$ represents rectilinear propagation orthogonal to the singular plane. Once $x(t)$ is obtained, we can use the second and third equations in (\ref{eq:x}) to get $y(t)$ and $z(t)$, respectively. In the case where $\Tilde{k}\neq 0$, they read 
\begin{equation}
\left\{\begin{array}{rl}
       y(t) \\
       z(t)\\
    \end{array}\right\}=\left\{\begin{array}{rl}
       y_{0} \\
       z_{0}\\
\end{array}\right\}\pm\left\{\begin{array}{rl}
       k_{y} \\
       k_{z}\\
\end{array}\right\}\log\left|\tan(|\Tilde{k}|\, t+t_{0}/2)\right|\,,  \label{eq:sol_y}
\end{equation}
and, of course, $y,z=const.$ when $\Tilde{k}=0$. Additionally, from the dispersion relation $\mathrm{H}_{am}=0$, we have
\begin{equation} 
k_x(x)=\pm (x^2-\Tilde{k})^{1/2}\,. 
\label{elkog}
\end{equation}
It is quite clear that light rays never reach the singular plane, not even when they propagate right into it, see Eq. (\ref{eq:sol_x_k_0}); they will take an infinite time to hit the singularity. This is the manifestation in the lab (material medium) of the fact that null geodesics are complete, i.e., they are defined for all values of the affine parameter. 
\begin{figure}
    \centering
\includegraphics[width=.85\linewidth]{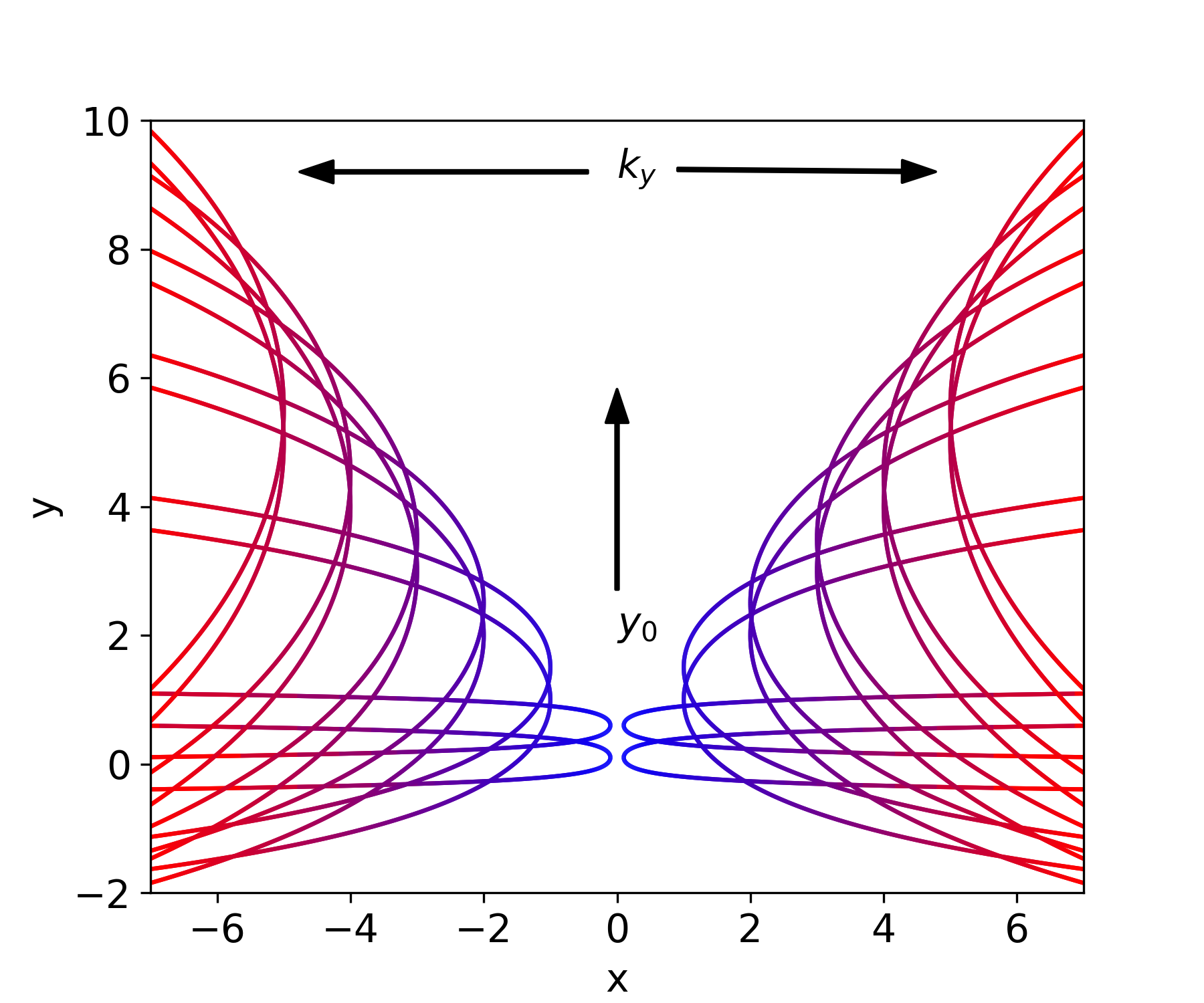}
    \caption{A family of GO trajectories (equivalent to null geodesics) as come from Eqs.~(\ref{eq:sol_x})--(\ref{eq:sol_y}).
    }
    \label{fig:x_y_analogo}
\end{figure}

A family of light-rays are depicted in Fig.~\ref{fig:x_y_analogo} for $k_z=0$, therefore in the $x$--$y$ plane. The curves correspond to values of $y_0$ and $k_y$ varying according to the arrows shown in the picture. In particular, $k_y$ takes the values $0.1, 1, 2, 3, 4, 5$. It can be seen how, as $|\tilde{k}|=|k_y| \rightarrow 0$, the trajectories asymptotically approach the singular plane without ever reaching it. The shading in the curves represents increasing values of $|\bar{k}(\bar{x})|$, from blue to red.


\section{Beyond GO: an extension thereof} As mentioned, GO approximation is by no means reliable in the strong-field regime, because the fields are not slowly varying there in comparison with the scale $k_{0}$. This is why we will proceed further by considering waves beyond the GO limit. If we particularize the full equations (\ref{eq:rot_E_cuasi_planas}) and (\ref{eq:rot_H_cuasi_planas}) for the material medium described by (\ref{eq:K}), we obtain
\begin{align}
&i\,k_{0}^{-1}\,\bar{\nabla}\times\bar{E}=\bar{\nabla}(\bar{k}\cdot\bar{x})\times\bar{E}-|x|\bar{H}\,, \label{eq:rot_E_cuasi_planas_met_sencilla}\\
&i\,k_{0}^{-1}\,\bar{\nabla}\times\bar{H}=\bar{\nabla}(\bar{k}\cdot\bar{x})\times\bar{H}+|x| \bar{E}\label{eq:rot_H_cuasi_planas_met_sencilla}\,.
\end{align}
As an example, let us consider that the electric field has only $z$-component, which depends solely on $x$ and $y$, namely $\bar{E}(\bar{x})=E^z(x,y)\hat{z}$. This greatly simplifies the analysis, because, after plugging $\bar{H}$ as comes from Eq. (\ref{eq:rot_E_cuasi_planas_met_sencilla}) into Eq. (\ref{eq:rot_H_cuasi_planas_met_sencilla}), the $x$ and $y$ components of the latter are identically zero. We are left with the $z$-component alone, which the reader can find written in Eq. (\ref{ectot}) of the Appendix in terms of the function $F(x)=x\,\partial_{x} k_{x}+k_x$. 

Now, we would like to find planar solutions to (\ref{ectot}), with a wave vector of the form (\ref{elkog}),
which is the form that $\bar{k}$ has in the $k_{x}$--$k_{y}$ plane when the GO limit is considered. We can then plug $F(x)=x\,\partial_{x} k_{x}+k_x=x^2/k_{x}+k_{x}$, and $\partial_{x}F=x\,(3-x^2/{k_{x}}^2)/k_{x}$ in (\ref{ectot}) to obtain  
\begin{align}
    &\left[i\left(\frac{k_x}{x}-\frac{2\,x}{k_x}+\frac{x^3}{{k_x}^3}\right)+2\,k_0x^2+\frac{k_0 x^4}{{k_x}^2}\right]E^z \nonumber \\
    &-i\left(\frac{i}{k_0 x}+\frac{2\,x^2}{k_x}+2\,k_x\right)\partial_{x} E^z \\
    &-2\,i\,k_y\partial_{y} E^z-k_0^{-1}\left(\partial^2_{x} E^z+\partial^2_{y} E^z\right)=0\,. \nonumber
\end{align}
This equation is separable by writing $E^z(x,y)=X(x)\,Y(y)$, giving rise to the two differential equations
\begin{align}
    Y''&=k_0\,c^2\, Y-2\,i\,k_0\, k_y\,Y'\,,\label{eq:Y_y_cte_0}\\
    \frac{X''}{k_0}&=\left[i\left(\frac{k_x}{x}-\frac{2\,x}{k_x}+\frac{x^3}{{k_x}^3}\right)+2\,k_0\,x^2+\frac{k_0x^4}{{k_x}^2}-c^2\right]X \nonumber\\
    &-i\left(\frac{i}{k_0 x}+\frac{2\,x^2}{k_x}+2\,k_x\right)X'\,, \label{eq:X_x_cte_0}
\end{align}
where $c^2$ is the separation constant, and it is worth remembering that $k_{x}$ is given in Eq. (\ref{elkog}).
We will now divide the analysis into two scenarios, based on the possible values of the separation constant.

\bigskip

\textbf{(a) Case $\mathbf{c^2+k_y^2\neq 0}$.} The solutions are 
\begin{align}
Y=&\,c_1\,\operatorname{e}^{-i\, k^{+}\, y}+c_2\,\operatorname{e}^{-i\,k^{-}\,y} \,,\label{eq:sol_Y_k2_dist_0}\\
X=&\operatorname{e}^{\frac{i\,k_0}{2}\,x\left(x-2k_{x}\right)} 
 \Big[c_3\,U\left(A,0,B\right) 
 \left.+c_4\,L_{(-A)}^{(-1)}(B)\right], \label{eq:sol_X_k2_dist_0} 
\end{align}
where $k^{\pm}\doteq k_0[\, k_{y}\pm({k_y}^2-c^2)^{1/2}]$, $A=i\,k_0\,\left({k_y}^2-c^2\right)/4$, $B=B(x)= -i\,k_0\, x^2$ and $c_{i}$ $i=1,2,3,4$ are integration constants in the reals (the separation constant was redefined according to $ c^2/k_{0}\rightarrow c^2$). The new functions arising in Eq. (\ref{eq:sol_X_k2_dist_0}) are the confluent hypergeometric function of the second kind, $U(a,b,z)$, and the generalized Laguerre polynomials, $L_{(\beta)}^{(\alpha)}(x)$, see the Appendix for the corresponding definitions and asymptotic properties of these two functions. 
Note that the hypergeometric function $U$ takes a nonzero, finite value at $x=0$, while the generalized Laguerre polynomial $L_{(-A)}^{(-1)}$ vanishes there, see Eqs. (\ref{or1}) and (\ref{or2}). Therefore, the function $X(x)$ given by (\ref{eq:sol_X_k2_dist_0}) is regular at $x=0$. Moreover, $  \lim_{{x\to 0}}E^z=c_3 Y(y )/A\,\Gamma\left(A\right)$, see (\ref{sor2}). Hence, the electric field is also regular at $x=0$, regardless of the value of $y$.

On the other hand, using in Eq. (\ref{eq:rot_E_cuasi_planas_met_sencilla}) that $\bar{E}(\bar{x})=E^z(x,y)\hat{z}$, the nonvanishing magnetic field components are 
\begin{align}
    H^x=|x|^{-1}&\left(k_y E^z-i\,{k_0}^{-1}\partial_{y} E^z\right)\,,\\
    H^y=|x|^{-1}&\left(-F E_0^z 
    +i\,{k_0}^{-1}\partial_{x} E^z\right)\,, 
\end{align}
which, in terms of the functions $X(x)$ and $Y(y)$, result 
\begin{align}
&H^x =X\,|x|^{-1}\left[k_y Y-i\,Y^{\prime}/k_0\right]\,,\label{hx2}\\
&H^y=-Y\,|x|^{-1}\left[\left(k_{x}+{k_{x}}^{-1}x^2\right)X-i\, X^{\prime}/k_0\right]\,.  \label{hy2}
\end{align}
In view that $k_{x}$ is given in Eq.~(\ref{elkog}) and due to the asymptotic behavior of the function $X$ as $x\rightarrow0$, see Eq.~(\ref{sor2}), we see that near $x=0$ the magnetic field has the structure 
\begin{align}
&H^x =\left[ \frac{c_{3}}{|x|A\Gamma(A)}+ H_{0}^x +\mathcal{O}(x)\right]\left[k_y Y-i\,Y^{\prime}/k_0\right],\label{hx22}\\
&H^y=-\left[H_{0}^y+\mathcal{O}(x)  \right]\,Y,  \label{hy22}
\end{align}
where the two constants $H_{0}^x$ and $H_{0}^y$ read
\begin{align}
H^x_{0}=& \pm\frac{c_{3}\,k_{0}|k_{y}| \operatorname{sgn}(x)}{A \,\Gamma(A)}\,, \label{hx22c}\\
H^y_{0} =&\frac{i\,c_{3}\,k_{0}{k_{y}}^2\operatorname{sgn}(x)}{A \,\Gamma(A)}+\frac{2\operatorname{sgn}(x)}{i\,k_{0} }\bigg[c_{3}U_{0}+i c_{4}k_{0}\notag \\
&+\frac{i\,k_0\,c_3}{2\,A\,\Gamma(A)}\left(1-i\,k_0\,{k_y}^2\right)\bigg] \,.  \label{hy22c}
\end{align}
Then, the important fact is that for $c_3=0$ the magnetic field is also regular at $x=0$. Actually, $H^x$ is zero there and $\lim_{{x\to 0^{\pm}}}H^y(x,y)=-2\,c_{4}\operatorname{sgn}(x)Y(y)$, so $H^y$ is bounded but discontinuous. From now on we will focus on this choice of $c_{3}$. 

\begin{figure}
    \centering
\includegraphics[width=.95\linewidth]{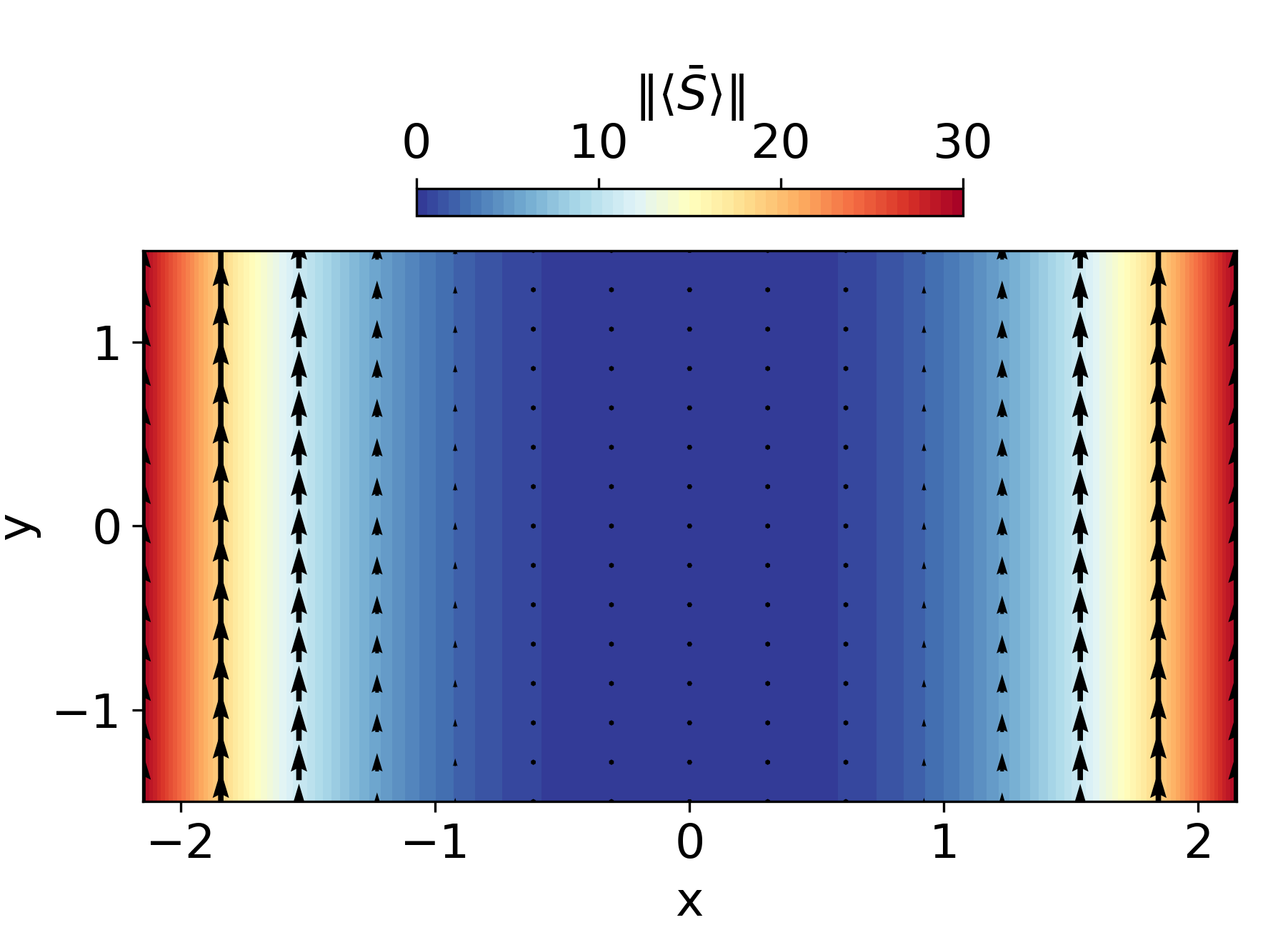}
    \caption{Time averaged Poynting vector in the vicinity of the singular plane $x=0$, as given by Eq.~(\ref{els1f_imky0}) for real $\kappa$ and $c_2 > c_1$.}  
    \label{fig:S_quiver}
\end{figure}

As a consequence of the regularity of the EM field at $x=0$, and of the zero value of the electric component there, the time-averaged power density given by the Poynting vector vanishes at $x=0$. Precisely, in a period $T$ we have
\begin{align}
    \langle \bar{S} \rangle &=T^{-1}\int_0^T \bar{S}(t)\,dt= \operatorname{Re}\left[\bar{\mathcal{E}}\times\bar{\mathcal{H}}^{*} \right]/2\nonumber \\
       &= \operatorname{Re}\left[-E^z \left(H^y \right)^{*} \hat{x} +E^z \left(H^x\right)^*\hat{y}\right]/2\label{els1},
  \end{align}
which turns, as $x\rightarrow 0$, into
\begin{align}
& \langle \bar{S} \rangle \approx\hat{y}\, {c_{4}}^2\,{k_0}^2 |x|^3\Big[k_y |Y|^2 +k_{0}^{-1}\operatorname{Re}(iY (Y^{\prime})^*)\Big]/2\,,
\end{align}
where we considered $\operatorname{Im}(k_y)=0$ and the asymptotic expressions for $E^z$, $H^x$ and $H^y$ were used, see Eqs. (\ref{sor2}), (\ref{hx22}) and (\ref{hy22}), respectively. Employing  (\ref{eq:sol_Y_k2_dist_0}), this results
\begin{align} 
     \langle \bar{S} \rangle  \approx \hat{y}\,{c_{4}}^2\,k_0 |x|^3 \operatorname{Re}(\kappa) (c_2^2 - c_1^2)/2, \label{els1f_imky0}
\end{align}
where $\kappa=k_{0}({k_{y}}^2-c^2)^{1/2}$, since $\kappa$ is either real or pure imaginary. Notice that, provided $\kappa\in\mathbb{R}$ and
$c_1 \neq c_2$, there is a net power flux always directed along $\hat{y}$ (upwards or downwards depending on the sign of $c_2^2 - c_1^2$). This is consistent with the fact that the wave vector is the one corresponding to the GO limit, where a ray approaches the singularity obliquely and then moves away after being repelled (see Fig. \ref{fig:x_y_analogo} for the light trajectories in GO). In Fig.~\ref{fig:S_quiver} we have depicted the time-averaged Poynting vector for a case in which $c_2 > c_1$ and $\kappa\in \mathbb{R}$. In this regard, it would be fair to say that the singularity is acting as a perfect conducting plane, consistent with a zero tangential electric field $E^z$ and a zero normal magnetic field $H^x$ at the singularity, see Eq. (\ref{hx22}). This is valid for any wave number $k_{0}$, thus constituting a result which applies beyond the realm of GO. This simple example shows how this naked singularity sorts things out so as to eject the incoming EM radiation.
\bigskip

\textbf{(b) Case $\mathbf{c=k_y=0}$.} In this particular instance the wave motion is one-dimensional because $\bar{k}=\pm |x|\hat{x}$, and the wave is propagating right into, or out of the singularity. Eqs. (\ref{eq:Y_y_cte_0}) and (\ref{eq:X_x_cte_0}) acquire the form
\begin{align}
    Y''=&\,0\,,\\
   X''=& 3\,{k_0}^2\,x^2 X-i\left(i\,x^{-1}\pm4k_{0}|x|\right)X'\,.
\end{align}
The presence of the $\pm$ sign in front of $|x|$, which defines two branches, leads to different scenarios according to the sign of the vector $\bar{k}$ on either side of the singular plane. 

The intervening solutions are
\begin{align}
Y(y)=&c_1\,y+c_2\,, \label{eq:sol_Y_y_k2_0}\\
X_1(x)=&\operatorname{e}^{-\frac{3}{2}i\,k_0\,x^2}   \left(c_3\operatorname{e}^{i\,k_0\,x^2}+c_4\right),\label{elx01}\\
X_2(x)=&\operatorname{e}^{\frac{3}{2}i\,k_0\,x^2} 
\left(c_3\operatorname{e}^{-i\,k_0\,x^2}+c_4\right),\label{elx02}
\end{align}
where $c_{i}$, $i:1,2,3,4$, are real integration constants, not related to the previous ones. The solution $X_1(x)$ corresponds to the plus sign for $\bar{k}$ when $x>0$, or the minus sign for $\bar{k}$ when $x<0$. Conversely, $X_2(x)$ arises from the plus sign for $\bar{k}$ when $x<0$, or the minus sign for $\bar{k}$ when $x>0$. Eqs. (\ref{elx01}) and (\ref{elx02}) should be properly combined to account for the two branches. Precisely, we have the following four schemes:
\begin{enumerate}   
\item $X_1(x)$, $\forall x$ ($\bar{k}=x\,\hat{x}$, $\forall x$),
\item $X_2(x)$, $\forall x$ ($\bar{k}=-x\,\hat{x}$, $\forall x$),
\item $X_1(x)$, $x\geq0$,\, \text{and}\, $X_2(x)$, $x\leq0$ ($\bar{k}=|x|\hat{x}, \,\,\forall x$),
\item $X_2(x)$, $x\geq0$,\, \text{and}\, $X_1(x)$, $x\leq0$ ($\bar{k}=-|x|\hat{x}, \,\,\forall x$).
\end{enumerate}
In any event, the electric field $E^z= X(x) Y(y)$ is regular and nonzero at $x=0$. Moreover, the non-vanishing components of $\bar{H}$, obtained from (\ref{hx2}) and (\ref{hy2}) with $k_{x}=\pm |x|$ and $k_{y}=0$, are given by
\begin{align}
    H^x&=-i\, c_1\,{k_0}^{-1}\,|x|^{-1} X\label{magxk1},\\
    H^y&=Y \left[\mp\,2\,X +i\,{k_0}^{-1}\,|x|^{-1}X'  \right] \label{magxk2} \,.
\end{align}
In view of the nonzero character of the function $X$ at $x=0$, we see that $H^x$ diverges there if $c_1\neq0 $. In contrast, $H^y$ remains regular at $x=0$. Thus, if $c_{1}=0$, we have a TEM wave with finite field amplitude at the singularity, carrying a time-averaged power density (see Eq. (\ref{els1}))
\begin{align}
    \langle \bar{S} \rangle=-\hat{x} \operatorname{Re}\left[E^z \left(H^y \right)^{*} \right]/2, 
 \end{align}     
which, after using (\ref{magxk2}), becomes
\begin{align}
    \langle \bar{S} \rangle=c_{2}^2 \left[\pm|X|^2+ {k_{0}}^{-1} |x|^{-1}   \operatorname{Re}\left[i X (X')^{*}\right]/2\right] \hat{x}.   
    \label{poynting}
 \end{align}  
This can be easily computed taking into account the scheme outlined in the points (1)-(4) above, and using (\ref{elx01}) and (\ref{elx02}). At once we get
\begin{equation}
    |X_{1,2}|^2={c_3}^2+{c_4}^2+2\,c_3\,c_4 \cos{\left(k_0 x^2\right)},\notag
\end{equation}
and, on the other hand,
\begin{align}
    \operatorname{Re}\left[i
    X (X')^{*}\right]=\pm k_0 x\left[{c_3}^2+3{c_4}^2+4 c_3 c_4\cos{\left(k_0 x^2\right)}\right],\notag
\end{align}
where the $-$ sign corresponds to $X_{1}$, and the $+$ sign to $X_{2}$. Without loss of generality, because $c_{2}\neq0$, we set ${c_{2}}^2/2=1$, which is equivalent to redefine $c_{3}$ and $c_{4}$. We compute (\ref{poynting}) for the different situations involved; for instance, scheme 1 gives (the same for scheme 2 with a global minus sign)
\begin{align}\label{outin}
\langle \bar{S} \rangle_{out/in}&=\hat{x}\left\{\begin{array}{rl} {c_3}^2-{c_4}^2\,,&x>0 \\
      {c_4}^2-{c_3}^2\,,&x<0\,. 
      \end{array}
      \right.
\end{align}
In this case we have a discontinuous power flux at the singular plane. This is because the energy is either flowing into the singularity from both sides, or out of it, according to the sign of ${c_3}^2-{c_4}^2$. The choice $c_3^2=c_4^2$ results in no energy flux whatsoever. This is because the EM field becomes a standing wave on either side, carrying no net energy. However, if $c_3^2\neq c_4^2$, and unlike in the previously considered example, the electric field is not zero at $x=0$, then, the singular plane is not reflecting the waves. Rather, we can interpret the situation as if the singularity were creating (emitting) or absorbing (receiving) the waves without causing any harm on them. 

Finally, we also have the following possibility
\begin{equation}\label{cross}
\langle \bar{S} \rangle_{cross}=\hat{x}\left\{\begin{array}{rcl}
      {c_3}^2-{c_4}^2\,,&\forall x &\text{(scheme 3)}\\
      {c_4}^2-{c_3}^2\,,&\forall x &\text{(scheme 4),}
      \end{array}
      \right.
\end{equation}
which, surprisingly enough, represents a constant power flux moving from either side of the space to the other, through the singularity. 


\section{Final comments} 
Our austere example just pretends to be a contribution in order to show that physics in close proximity to a strong curvature singularity is not necessarily that remediless, destructive thing we thought it was, although most of the time it certainly is, as in the solutions we have discarded precisely for that reason by setting $c_{3}=0$ in Eqs. (\ref{hx22}) and (\ref{hy22}), or $c_{1}=0$ in Eq. (\ref{magxk1}). By means of Eqs. (\ref{els1f_imky0}), (\ref{outin}) and (\ref{cross}) we were able to pronounce sharp, concrete answers to the questions raised in the introduction. Our work reinforces recent investigations aiming to explain trans-singularity, pre-Big Bang physics, see, e.g. \cite{singularitycrossing1}-\cite{singularitycrossing2}. This area of research is not only growing from a theoretical point of view in relation to the development, for instance, of Conformal Cyclic Cosmology \cite{Penrosec}, \cite{ccc}, but also from an observational perspective with regard to the uncanny discovery of galaxies at surprisingly high redshifts, consult \cite{z13} and \cite{z14}.

\bigskip

\textbf{\emph{Acknowledgements}}. FF and SMH are members of \emph{Carrera del Investigador Cient\'{i}fico} (CONICET). Their work is supported by CONICET and Instituto Balseiro (UNCUYO). JMP is a PhD student supported by CONICET.  
\appendix \label{ap1}
\section{}
Under the ansatz $\bar{E}(\bar{x})=E^z(x,y)\hat{z}$, the combination of $\bar{H}$ as comes from Eq. (\ref{eq:rot_E_cuasi_planas_met_sencilla}) with Eq. (\ref{eq:rot_H_cuasi_planas_met_sencilla}), leads us to the sole equation
\begin{widetext}
\begin{align}\label{ectot}
   \left(ix^{-1} F- i\,\partial_{x} F+ k_0 {k_y}^2+k_0 F^2-k_0 x^2\right)E^z+  \left({k_0}^{-1}\,x^{-1}-2i F  \right)\partial_{x} E^z-2\,i\, k_{y}\partial_{y} E^z- {k_0}^{-1} \left(\partial^2_{x} E^z+\partial^2_{y} E^z\right)=0\,,
\end{align}
\end{widetext}
which is the starting point of the subsequent analysis.

On the other hand, let us proceed now to briefly comment on the functions taking part in Eq.  (\ref{eq:sol_X_k2_dist_0}). For details we refer the reader to, e.g., \cite{hiper}. The confluent hypergeometric function of the second kind
\begin{align}
   &U(a,0,z)=\frac{1}{\Gamma(a+1)}\Big[1+a\,z\, Log(z)\,  _{\textbf{1}}F_{\textbf{1}}(a+1,2,z)\notag\\
  &-\sum_{k=1}^\infty\frac{(a)_s}{k! (k-1)!}\left(2\psi(k)-\psi(a+k)+k^{-1}\right)z^k\Big]\,,
\end{align}
is defined in terms of the Gamma function
\begin{align}
    &\Gamma(z)=\int_0^\infty t^{z-1} \operatorname{e}^{-t} dt\,,\,\,\,\,Re(z)>0,
\end{align}
the di-gamma function $\psi(z)=\Gamma^{\prime}(z)/\Gamma(z)$, and the generalized hypergeometric function
\begin{equation}
    _{\textbf{1}}F_{\textbf{1}}(b,c,z)=
    \sum_{s=0}^\infty\frac{(b)_s}{(c)_s}\frac{z^s}{s!},\notag
\end{equation}
which is constructed in terms of the rising factorial
\begin{align}
    &(b)_0=1\,,\notag\\
    &(b)_s=b(b+1)(b+2)\dots(b+s-1)\hspace{0.5cm}(s\geq1)\,.\notag
\end{align}
In (\ref{eq:sol_X_k2_dist_0}) also appear the generalized Laguerre polynomials, which admit the integral representation
\begin{align}
    &L_{(\beta)}^{(\alpha)}(x)=\frac{1}{2\,\pi\,i}\oint_C\frac{\operatorname{e}^{-x\,t/(1-t)}}{(1-t)^{\alpha+1}t^{\beta+1}}dt\,.
\label{eq:pol_generalizado_laguerre}
\end{align}
Here the line integral is performed on any loop enclosing the origin, but not the point $t=1$. The structure of the two functions near the the origin is given by
\begin{align}
    & U(A,0,-i k_0 x^2)=\frac{1}{A\,\Gamma(A)}+ U_{0}\,x^2+\mathcal{O}(x^4)\,,\label{or1}\\
    & L_{(-A)}^{(-1)}\left(-i k_0 x^2\right)=i\,k_0\,x^2+ \mathcal{O}(x^4)\,\label{or2},
\end{align}
where $U_{0}=U_0(A)= ik_{0}\left[1-2\gamma-\psi(1+A)\right]/\Gamma(A)$ and $\gamma$ is the Euler constant. Consequently, at second order, the function $X(x)$ near $x=0$, after using (\ref{eq:sol_X_k2_dist_0}) and (\ref{or1})-(\ref{or2}), results 
\begin{align}
  X=&\frac{c_{3}}{ A\,\Gamma(A)} \pm\frac{k_{0}|k_{y}|\,c_{3}\,x}{ A\,\Gamma(A)}+\bigg[c_{3}U_{0}+i c_{4}k_{0}\notag \\
  &+\frac{i\,k_0\,c_3}{2\,A\,\Gamma(A)}\left(1-i\,k_0\,{k_y}^2\right)\bigg]\,x^2+\mathcal{O}(x^3) \label{sor2},
\end{align}
where $\pm$ comes from the fact that $k_{x}=\pm(x^2-{k_{y}}^2)^{1/2}$. This expression, and its derivative, is used in Eqs. (\ref{hx2}) and (\ref{hy2}) to obtain (\ref{hx22}) and (\ref{hy22}).


\begin{thebibliography}{99}
\bibitem{Penrose} R. Penrose, Nuovo Cimento \textbf{1} (1969) 252.
\bibitem{Penrose1} R. Penrose, Ann. New York Acad. of Sci. \textbf{224} (1973) 125.
\bibitem{Penrose2} R. Penrose, \emph{Singularities of space-time}, in Theoretical Principles in Astrophysics and Relativity, N.R. Levobitz, W.H. Reid and P.O. Vandervoort (eds.), University of Chicago Press (1981).
\bibitem{Harada} T. Harada, Pramana \textbf{63} (2004) 741.
\bibitem{Ong} Y. C. Ong,  Int. J. of Mod. Phys. \textbf{A35} (2020) 2030007.
\bibitem{Vir1} K.S. Virbhadra, D. Narasimha and S.M. Chitre, Astron. Astrophys. \textbf{337} (1998) 1.
\bibitem{Vir2} K.S. Virbhadra and G.F.R. Ellis, Phys. Rev. \textbf{D65} (2002) 103004.
\bibitem{Vir3} K.S. Virbhadra and C.R. Keeton, Phys. Rev. \textbf{D77} (2008) 124014.
\bibitem{nos2} F. Fiorini and J. M. Paez, Phys. Rev. \textbf{D111} (2025) 024063.
\bibitem{Newman} A. Janis, E. Newman and J. Winicour, Phys. Rev. Lett. \textbf{20} (1968) 878.
\bibitem{Maluf} J. W. Maluf, Gen. Rel. Grav. \textbf{46} (2014) 1734.
\bibitem{Glavan} D. Glavan and C. Lin, Phys. Rev. Lett. \textbf{124} (2020) 
081301.
\bibitem{nos3} C. G. Boehmer and F. Fiorini, Class. Quantum Grav. \textbf{37} (2020) 185002; ibid, \textbf{36} (2019) 12LT01.
\bibitem{Schmidt} B. G. Schmidt, Gen. Rel. Grav. \textbf{1} (1971) 269.
\bibitem{HE} S. W. Hawking and G. F. R. Ellis, \textit{The Large Scale Structure of Spacetime} (1973), Cambridge University Press.
\bibitem{Tamm} I. E. Tamm, J. Russ. Phys.-Chem. Soc., Phys. Section. \textbf{56} (1924) 248.
\bibitem{Pleb} J. Plebanski, Phys. Rev. \textbf{118} (1960) 1396.
\bibitem{leon-phil} U. Leonhardt and T. G. Philbin, \textit{Geometry and Light. The Science of Invisibility} (2010), Dover Publications Inc.
\bibitem{leon-phi2} U. Leonhardt and T. G. Philbin,  Prog. Opt. \textbf{53} (2009) 69.
\bibitem{Oleg} I. Fernández-Núñez and O. Bulashenko, Phys. Lett. \textbf{A380} (2016) 1.
\bibitem{Turner} R. A. Tinguely and A. P. Turner, Comm. Phys. \textbf{3} (2020) 120.
\bibitem{Gaitas} E. Falcón-Gómez \emph{et al}, Phys. Rev. \textbf{D110}  (2024) 084002. 
\bibitem{RAD} J. Drori \emph{et al}, Phys. Rev. Lett. \textbf{122} (2019) 010404. 
\bibitem{Hamop} V. Perlick, \textit{Ray optics, Fermat´s Principle, and Applications to General Relativity} (2000), Springer.
\bibitem{Mackay1} T. G. Mackay and A. Lakhtakia, \textit{Electromagnetic Anisotropy and Bianisotropy: A Field Guide} (2019) Sec. ed., World Scientific.
\bibitem{nos} F. Fiorini, S. Hernández and E. Losada, Phys. Rev. \textbf{D104} (2021) 124009. 
\bibitem{hiper} G. E. Andrews, R. Askey and R. Roy, \textit{Special Functions}, in Encyclopedia of Mathematics and its Applications (1999), Cambridge University Press. 
\bibitem{singularitycrossing1} F. D'Ambrosio and C. Rovelli, Class. Quantum Grav. \textbf{35} (2018) 215010.
\bibitem{singularitycrossing12} M. Adamo and F. Mercati, Phys. Rev. \textbf{D110} (2024) 104033.  
\bibitem{singularitycrossing2} A. Rod Gover, J. Kopiński and A. Waldron, Phys. Rev. Lett. \textbf{133} (2024) 011401.
\bibitem{Penrosec} R. Penrose, \textit{Cycles of Time: An Extraordinary New View
of the Universe} (2010), Bodley Head.
\bibitem{ccc} M. Eckstein, Gen. Rel. Grav. \textbf{55} (2023) 26.
\bibitem{z13} R. P. Naidu \emph{et al}, ApJL \textbf{940} (2022) L14.
\bibitem{z14} S. Carniani \emph{et al}, Nature \textbf{633} (2024) 318.
\end{thebibliography}
\end{document}